\documentclass[useAMS,usenatbib]{mn2e}

\usepackage{graphicx,amssymb,amsmath}

\title[Realistic tidal field in Monte Carlo simulations]{Treatment of realistic
tidal field in Monte Carlo simulations of star
clusters}
\author[Sollima et al.]{A. Sollima$^{1}$\thanks{E-mail:
antonio.sollima@oabo.inaf.it}, A. Mastrobuono Battisti$^{2}$\\
$^{1}$ INAF Osservatorio Astronomico di Bologna, via Ranzani 1, Bologna, 40127,
Italy\\
$^{2}$ Physics Department, Technion - Israel Institute of Technology, Haifa,
32000, Israel}
\begin{document}


\pagerange{\pageref{firstpage}--\pageref{lastpage}} \pubyear{2013}

\maketitle

\label{firstpage}

\begin{abstract}
We present a new implementation of the Monte Carlo method to simulate the
evolution of star clusters. The major improvement with respect to the previously
developed codes is the treatment of the external tidal field taking into account for
both the loss of stars from the cluster boundary and the disk/bulge shocks. 
We provide recipes to handle with eccentric orbits in complex galactic 
potentials. The first calculations for stellar systems containing 21000 and
42000 equal-mass particles show good agreement with 
direct N-body simulations in terms of the evolution of both the enclosed mass 
and the Lagrangian radii provided that the mass-loss rate does not exceed a
critical value.
\end{abstract}


\section{Introduction}
\label{intro_sec}

The dynamical evolution of dense star clusters is a problem of fundamental 
importance in theoretical astrophysics. Star clusters like open and globular
clusters are among the simplest stellar systems: they are spherical, they
contain no dust to confuse the observations and they appear to have no dark 
matter. Moreover, they are dynamically old: a typical star in a globular cluster
has completed some $10^{4}$ orbits since the cluster was formed and processes
like gravothermal collapse and two-body relaxation occur on timescales
comparable with their ages. Thus, they provide the best physical realization of
the gravitational N-body problem i.e. to understand the evolution of a system of
N point masses interacting only by gravitational forces.
In spite of the many advances made in the recent past, many aspects of the problem 
have remained unresolved like the production of exotic objects (Ferraro et al.
2012), the importance of tidal-shocks in the long term evolution and survival 
of star clusters in the Galaxy (Gnedin, Lee \& Ostriker 1999) and the ability 
to retain dark remnants (Morscher et al. 2013; Sippel \&
Hurley 2013). The most direct approach to the simulation of star clusters is
through N-body simulations. In these kind of studies the gravitational forces of
stars are directly computed and any additional ingredient like e.g.
binaries, tidal field, stellar evolution, etc. can be easily incorporated.
For this reason, in many cases N-body simulations represent the unique tool to 
face with complex topics within the gravitational N-body problem.
However, several processes involved in the dynamical evolution of a star
cluster occur on different timescales, so that a direct scaling
of the result of an N-body simulation to larger number of particles is not
possible (Baumgardt 2001).
Although the GRAPE series of special-purpose computers is steadily 
increasing in performance and the development of Graphic Processing Units (GPUs)
computing, direct N-body simulation of the evolution of clusters with more 
than a few percent binaries and a moderate number of stars ($10^{5}$) is still 
computationally expensive, with computational timescales of the order of 
months. Until now only open clusters such as M67 and the Arches cluster 
(Hurley et al. 2005; Harfst, Portegies Zwart \& Stolte 2010) and loosely bound 
globular cluster objects such as Palomar 4 and Palomar 14 (Zonoozi et al. 
2011, 2014) have been modelled at the necessary level of sophistication.

Alternative numerical methods to simulate the evolution of star clusters have 
been developed in past years including fluid models (Larson 1970; Angeletti \& 
Giannone 1977a,b), orbit-averaged Fokker-Planck methods (Cohn 1980; Takahashi 1995)
and Monte Carlo simulations (H{\'e}non 1971, hereafter H71; Giersz 1998; Joshi,
Rasio \& Portegies-Zwart 2000).
Monte Carlo methods can be regarded as a hybrid between direct N-body
integrations and numerical solutions of the Fokker-Planck equation. In this
approach the system is modelled as a sample of "superstars" i.e. a subsample of
stars sharing the same mass and integrals of motions. In spherical systems the
motion of each superstar depends only on its energy and angular momentum and on 
the cluster potential, this last quantity being a unique function of positions 
and masses of the superstars. Through an iterative algorithm it is therefore possible to 
follow the evolution of the system once suited perturbations to the integrals of
motion of the superstars are applied to account for the effect of two-body
interactions. Within the family of Monte Carlo methods two approaches can be
distinguished: {\it i)} the {\it orbit-following} method (also known as the "Princeton
method"; Spitzer \& Hart 1971) where the orbits of the superstars are directly computed and 
{\it ii)} the {\it orbit-averaged} method (the "Cornell method"; H71) where only
the energies and angular momenta of superstars are monitored. While
orbit-following methods are more suited to follow all those processes occurring on
the dynamical timescale (such as the evaporation of stars from
the cluster, the violent relaxation, the tidal shocks and the phase of post-core
collapse), orbit-averaged methods are computationally less expensive since
perturbations to the integrals of motions need to be computed at time-steps
that are short compared to the relaxation time and the time consuming 
integration of the orbits is not required. Moreover, it is 
particularly easy to add more complexity and realism to the  
simulations one layer at a time and they are particularly easy to be
parallelised. For these reasons, orbit-averaged Monte Carlo simulations have been employed by a
number of groups to study the dynamical evolution of globular clusters and the
dense centers of galaxies. In recent years a particular effort have been made by
these groups to include the effect of a mass spectrum (Giersz 2001), stellar 
evolution (Joshi, Nave \& Rasio 2001), three- and four-body interactions 
(Giersz \& Spurzem 2003; Fregeau \& Rasio 2003; Fregeau et al. 2007) and a simplified treatment 
of a tidal field (Giersz et al. 2013; Takahashi \& Baumgardt 2012). 

One of the most complex process to be
modelled in Monte Carlo simulations is the escape from a cluster in an
external tidal field. In fact, while stars can escape from an isolated cluster
only when they have positive energies, when the cluster moves within a tidal field the
effective potential felt by a cluster star is perturbed and has a maximum at a 
distance (called "tidal radius") which depends on the shape of the external field 
and on the orbital parameters of the cluster. If the star moves beyond this radius 
the gravitational attraction 
of the cluster will not balance the combined effect of the external
tidal field and the centrifugal
force and the star will escape from the cluster. For this reason the presence of
the external field accelerates the escape process and
consequently the whole structural evolution of the system (Spitzer 1987). Since the first
pioneering studies by H71, the effect of a steady external field has
been modelled by removing stars able to reach an apocenter larger that the tidal
radius. This last quantity was estimated considering the distance of the
Lagrangian point in the simple case of an
external potential produced by a point mass on a cluster moving on a circular
orbit. Even in this simple case, however, this criterion represents only a rough
approximation. Indeed, the presence of the external field breaks the spherical
symmetry of the effective potential and the size of the tidal radius depends on
the direction of the star motion. In practice, stars can escape from the cluster
only close to the direction of the Galactic center through the so-called "2nd
and 3rd Lagrangian points". Moreover, in real clusters, once a star reach
the energy required to escape, it needs several crossing times to reach the right
direction (the so-called "potential escapers") thus producing a delayed escape. 
Fukushige \& Heggie (2000) derived a
simple prescription to estimate the timescale of escape as a function of the
excess of energy with respect to the Lagrangian point energy level. On the basis
of this last result, Giersz et al. (2013) adopted a delayed escape criterion
which successfully account for this effect. The situation is more
complex when eccentric orbits are considered: in this case, the Hamiltonian is 
time-dependent and the tidal radius can be only instantaneously determined. 
In this situation, the aperture in the phase-space for a star to escape changes 
with time and stars can have only a limited amount of time to reach such an
aperture. Moreover, stars escaping from the cluster can be re-captured 
when the cluster expand during its motion away from the perigalacticon. To
further complicate the picture, it has been shown that stars with prograde and 
retrograde motion escape with
different efficencies (Read et al. 2006) and there are stars which permanently remain
bound to the cluster outside the tidal radius ("non-escapers"; Ross, Mennim \&
Heggie 1997).
Finally, the potential of the Milky Way cannot be realistically approximated
as a point mass but consists of many non-spherical components. In a 
non-spherical potential orbits are in general non-planar and rapid changes of
potential can produce compressive shocks which increase the kinetic energy
budget of cluster stars (Gnedin \& Ostriker 1997).

In this paper we present a new orbit-averaged Monte Carlo code able to simulate
the evolution of a star cluster moving on an eccentric orbit
within a realistic external potential. In Sect. 2 we describe the code and the
modification made to the original algorithm described by H71. In Sect. 3
the recipies to account for the effect of the external field are outlined. Sect.
4 in devoted to
the description of the set of simulations and their comparison with N-body
simulations. We summarize our results in Sect. 5. A detailed derivation of the
tidal radius and effective potential in complex potentials is provided in the
Appendix.

\section{Monte Carlo Technique} 
\label{mod_sec}

The code presented here is an updated version of the orbit-averaged Monte Carlo
method extensively described in H71 (see also Stodolkiewicz 1982, Giersz 1998 and Joshi et al. 2000). The
basic idea of this approach is to consider the cluster as a
sample of superstars characterized by mass ($m$), energy ($E$) and angular
momentum ($L$) per unit mass generating a spherical symmetric potential ($\phi$).
The evolution of the cluster is divided in time-steps ($\Delta t$) of variable 
duration. At each time-step the following steps are performed
\begin{enumerate}
\item The optimal time-step is determined (see eq. 10 by Joshi et al. 2000; see
also Sect. \ref{esc_sec});
\item A statistical realization of the cluster is performed by placing
the superstars at random positions along their orbits. Each star is placed at a
given distance from the cluster center according to the inverse of the
star velocity at that distance (see below; see also Sect. 7 of H71); 
\item The cluster potential
profile is evaluated according to the masses and positions of the
superstars (see below); 
\item The mechanical work made by the (internal+external) potential change on the suparstars is
calculated and corrections to the stars' energies are applied (see Sect.
\ref{shocks_sec}; see also Sect. 4 of Stodolkiewicz 1982);
\item Each superstar is
assumed to interact with its nearest neighbor producing a perturbation on its
energy and angular momentum (see Sect. 5 of H71). 
\item Stars satisfying the escape criterion (see Sect. \ref{esc_sec})
are removed from the simulation.
\end{enumerate}
The above steps are repeated until the end of the simulation.
A detailed description of the algorithms adopted to perform the above steps is
provided in H71, Giersz 1998 and Joshi et al. (2000) and will not be repeated here.
Below we describe only the modifications to their approaches regarding the way the
cluster potential is determined while in Sect. \ref{ext_sec} we extensively
describe the adopted escape criterion.

The most time-consuming step of the procedure outlined above is the distribution
of the superstars across
the cluster. This is done by extracting a variable $s$ (with $-1<s<1$) from the distribution 
$$g(s)=\frac{3 (1-s^{2})}{4 |v_{r}(s)|}(r_{min}+3 r_{max})$$
where $r_{min}$ and $r_{max}$ are the pericenter and apocenter of the star orbit
within the cluster potential and $v_{r}(s)$ is the radial component of the star
velocity at 
$$r\equiv\frac{1}{2}(r_{max}+r_{min})+\frac{1}{4}(r_{max}-r_{min})(3
s-s^{3})$$ 
The values of $r_{min}$ and
$r_{max}$ are first determined from the star energy and angular momentum and the
cluster potential at the two zeros of the function
\begin{equation}
v_{r}=\sqrt{2(E-\phi(r))-\frac{L^{2}}{r^{2}}}
\label{vr_eq}
\end{equation}
Then $s$ is extracted from $g(s)$ using the von Neumann rejection technique.
H71 demonstrated that the above algorithm ensures that the probability 
to extract a given position is proportional to the time spent by the star in that
position.
According to H71, the cluster potential at a given
distance from the cluster center $r$ can be
determined in a straightforward way from the positions and masses of the
superstars using the Poisson equation in its discrete integral form
\begin{equation}
\phi(r)=-G\left(\frac{1}{r}\sum_{i=1}^{k} m_{i}+\sum_{i=k+1}^{N}
\frac{m_{i}}{r_{i}}\right)
\label{pot1_eq}
\end{equation}
where $r_{i}$ and $m_{i}$ are the position and mass of the i-th superstar, $N$
is the total number of superstars and $k$ is the index such that
$r_{k}<r<r_{k+1}$.
This approach is efficient and adaptive by definition i.e. in the densest
regions of the cluster it provides a better sampling of the potential.
However, from a computational point of view, the above procedure is quite
expensive since it requires a cycle over the N superstars to find the index $k$
which must be repeated at least two times for each superstar to find 
$r_{min}$, $r_{max}$ and $v_{r}(s)$.
 
For this purpose we decided to calculate the cluster potential at the beginning
of each time-step on a grid of $M$ evenly spaced radial steps. The potential 
at the distance $r$ is then determined by linearly interpolating between
the two contiguous knots $k$ and $k+1$. In this case the index $k$ is
immediately found as $k=int(r/\Delta r+1)$. This simple modification speed up 
the entire process by a factor of ten.

We developed two independent methods to define the potential in the grid knots
which are used in different conditions. The first method (hereafter referred as
the {\it fast method}) is to use eq.
\ref{pot1_eq} defining at each time-step a step-size of the grid 
$\Delta r=0.02~r_{c}$ where
\begin{equation}
r_{c}=\frac{\sum_{i=1}^{N} m_{i} n_{i} r_{i}}{\sum_{i=1}^{N} m_{i} 
n_{i}}
\label{rc_eq}
\end{equation}
is the cluster core radius (Casertano \& Hut 1985) and $n_{i}$ is the superstar 
number density at $r_{i}$ calculated using the
50 nearest neighbor superstars.
This is the fastest method and provides a good accuracy for most of the cluster 
evolution when clusters with moderate concentrations and a 
large number of superstars are considered.
Unfortunately, something is lost in the above modification: in the advanced
stages of core collapse a large fraction of stars is contained within the
innermost radial bins. In this situation, the spatial resolution of the grid is
not adequate to follow the fast evolution of the cluster core and produces an
unrealistic delay of the core collapse.
Moreover, another drawback of this method (which is in common with the canonical
method adopted by H71) is that the potential
depends on the position of the superstars. As positions are randomly extracted,
it is possible that fluctuations in the potential are present when a small
number of superstars are considered. This effect is
particularly strong in the cluster core where the potential is determined by few
stars and can produce a "spurious relaxation". H71 have shown that
such an effect is negligible when a large number ($N>10^{3}$) of particles is used.
However, a pernicious effect is produced by such fluctuations when the 
correction for the mechanical work made by the potential (step (iv) of the 
above scheme; see Stodolkiewicz 1982) is considered. Indeed,
fluctuations are erroneously interpreted as real potential changes  
introducing spurious corrections in the stars' energies. On the long term, this
produces a drift in the total cluster energy accelerating the relaxation process even when a
number of superstars as large as $N\sim10^{4}$ is considered.
  
For this reason we developed another method (hereafter referred as the {\it
integral method}) to determine the cluster potential in
the grid knots. Consider a star with energy $E_{i}$ and angular momentum $L_{i}$ moving
in the potential $\phi$. When the star is in the j-th radial bin $r_{j}$ its 
radial component of the velocity will be given by eq. \ref{vr_eq} 
and its radial component of the acceleration will be
$$a_{r,i}(r_{j})=\left.\frac{d v_{r,i}}{d r}\right|_{r_{j}}=-\left.\frac{d\phi}{d
r}\right|_{r_{j}}-\frac{L_{i}^{2}}{r_{j}^{3}}$$
These quantities have been estimated using the potential and its derivative
calculated in the previous time-step.
In the approximation that the star moves with a uniformly accelerated motion,
the time spent to cross the interval ($r-\Delta r/2$, $r+\Delta r/2$) is
$$\Delta t_{ij}=\frac{\sqrt{v_{r,i}^{2}+a_{r,i}\Delta
r}-\sqrt{v_{r,i}^{2}-a_{r,i}\Delta r}}{a_{r,i}}$$
The probability to find the star in that interval will be
$$P_{ij}=\frac{2 \Delta t_{ij}}{T_{i}}$$
where 
\begin{equation}
T_{i}=2 \sum_{j=1}^{M}\Delta t_{ij}
\label{per_eq}
\end{equation}
is the orbital period of the superstar. 
The potential at each point of the grid can be then calculated through the relation
$$\phi (r_{j'})=-G\left(\frac{1}{r_{j'}}\sum_{j=1}^{k} \sum_{i=1}^{N} P_{ij} m_{i}+
\sum_{j=k+1}^{M} \frac{1}{r_{j}} \sum_{i=1}^{N} P_{ij} m_{i}\right)$$
where $k$ is the index such that $r_{k}<r_{j'}<r_{k+1}$.
This method has the advantage to depend only on the energies and angular 
momenta of the stars and not on the randomly extracted positions of the superstars.
In practice, it is equivalent to compute the potential from an infinite number 
of randomly extracted positions thus eliminating the problem of fluctuations.
Moreover, it allows to estimate the orbital period of each superstar 
(eq. \ref{per_eq}) which will be used in the escape algorithm (see Sect.
\ref{esc_sec}).
There are two main drawbacks of this methods: first, to determine $v_{r,i}$ and
$a_{r,i}$ it is necessary to use the potential profile calculated in the previous
time-step. For this reason, at odds with the {\it fast
method}, the potential can be computed only on grid of positions which is fixed
in time. Therefore, a small step-size is required from the beginning of the simulation to 
adequately follow the cluster evolution also in the advanced stages of core 
collapse. We found that a good sampling of the potential profile during the
entire cluster evolution is provided by the choice of $\Delta r=0.001 r_{c,0}$ 
where $r_{c,0}$ is the cluster core radius (see eq.
\ref{rc_eq}) at the beginning of the simulation. Second, the above method is computationally expensive and almost
all the improvement provided by the adoption of the evenly spaced grid is lost.

To optimize the speed of the simulation without losses of accuracy, during the
simulation we adopt the {\it fast method}
when $N(r<r_{c})>1000$, and switch to the {\it integral
method} when the above condition is not satisfied. 

\section{External Tidal Field}
\label{ext_sec}

In this section we describe the treatment of the external tidal field in our
Monte Carlo code. Here we considered two kinds of external field potential: the
first generated by a point mass galaxy with mass $M=10^{10}~M_{\odot}$ and the 
second by the analytical bulge+disk+halo potential defined in Johnston, Spergel
\& Hernquist (1995; hereafter J95). The orbit of the cluster within these potentials has been
computed starting from its orbital energy and z-component of the angular
momentum using a fourth-order Runge-Kutta algorithm
providing an accuracy in terms of energy conservation better than 
$\Delta E/E<10^{-10}$ during the entire evolution. The effect of dynamical
friction has been neglected since it is expected to be negligible on the star
cluster scale (Gnedin et al. 1999).

The presence of an external field imply an increase of star losses
because of two main processes: escape of stars through the cluster boundaries
and tidal shocks (this last process occurring only when the external potential
has a bulge/disk component). We discuss the algorithms to include the above processes in
the following sections.

\subsection{Escape from the tidal boundary}
\label{esc_sec}

\begin{figure}
 \includegraphics[width=8.7cm]{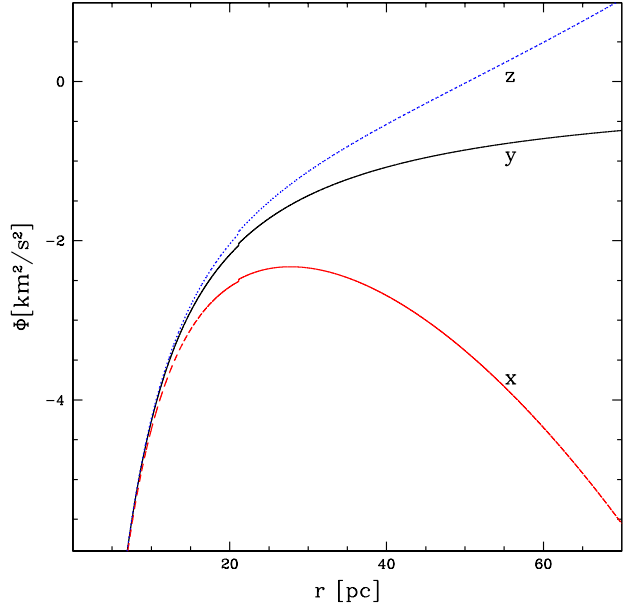}
 \caption{Effective potential of a King (1966) model with
 $M_{cl}=10^{4}M_{\odot}$ and $r_{c}=2~pc$ moving on a circular orbit
 at $r_{cl}=4~kpc$ around a point-mass galaxy of $M=10^{10}M_{\odot}$. The
 dashed, solid and dotted lines (red, black and blue in the online version of
 the paper) indicate the effective potential along the x-, y- and z-axis,
 respectively.}
\label{pot}
\end{figure}

As already introduced in Sect. \ref{intro_sec}, the presence of an external
tidal field imply the presence of a tidal cut in the cluster potential.
Stars with enough energy can cross the cluster tidal radius and evaporate from
the cluster on a timescale comparable to the star orbital period. 
A star orbiting around a cluster immersed in an external field 
feels an effective potential
given by the combination of the cluster potential ($\phi_{cl}$), the external
field potential ($\phi_{ext}$)
and a term linked to the angular motion of the cluster.
$$\phi_{eff}=\phi_{cl}+\phi_{ext}-\frac{1}{2}|{\bf \Omega} \times {\bf r}|^{2}$$
where {\bf $\Omega$} is the angular speed of the cluster and {\bf r} is the
position vector of the star in a reference frame centered on the
center of mass of the galaxy+cluster system rotating with angular speed ${\bf
\Omega}$.
Note that the Jacobi integral associated to the above effective potential is 
conserved only in the particular case of a
cluster moving on a circular orbit within a spherical potential. However, when
other orbits/potentials are considered, the timescale on which such an integral 
changes is longer than the dynamical time
of most cluster stars so that we can assume it instantaneously conserved. 
Consider a Cartesian reference system centered on the cluster with the x-axis pointed 
toward the galaxy center, the y-axis parallel to the galactic plane and directed 
toward the cluster rotation and the
z-axis perpendicular to the previous axes. 
The acceleration felt by the a star approaching the tidal radius $r_{t}$ will be
\begin{equation}
{\bf a}=-\frac{G M_{cl}{\bf r'}}{|{\bf r'}|^{3}}+({\bf r' \cdot \nabla}){\bf
\nabla \phi_{ext}}+{\bf a_{\Omega}}
\label{acc_eq}
\end{equation}
Where ${\bf r'}$is the position vector of the star, $M_{cl}$ is the cluster mass 
and ${\bf a_{\Omega}}$ is the acceleration due to the cluster angular motion.
The tidal radius is defined as the distance where the projection of the above 
acceleration on {\bf r'} is zero which corresponds to the radius at which the 
effective potential has a local maximum.
Note that, according to the above definition, both the tidal radius and the
effective potential depend on the
direction of the escape. It can be shown (see Appendix) that the shortest and
less energetic tidal
radius occurs in correspondence of the x-direction (i.e. the 2nd and 3rd
Lagrangian points) while in the z-direction
the effective potential is a growing function of the radius and no maxima
exist (see Fig. \ref{pot}). For this reason stars escape preferentially from the Lagrangian points.

To determine wether a star escapes from the cluster we defined three distinct
criteria. 
The first criterion we adopted is that the star energy and
angular momentum allow the motion across the tidal radius. For this purpose we
extracted two random numbers ($\eta_{1}$ and $\eta_{2}$) uniformly distributed between 
0 and 1 and defined the coefficients
\begin{eqnarray}
\tilde{x}&=&\eta_{1}\nonumber\\
\tilde{y}&=&\sqrt{1-\eta_{1}^{2}} sin(2\pi\eta_{2})\nonumber\\
\tilde{z}&=&\sqrt{1-\eta_{1}^{2}} cos(2\pi\eta_{2}).
\label{pos_eq}
\end{eqnarray}
These coefficient univocally define a direction of escape such that
${\bf r'}=\sum_{i=1}^{3} |{\bf r'}|\tilde{x}_{i}~{\bf \hat{e}_{i}}$. 
The tidal radius and
effective potential in the defined direction are then calculated (see Appendix)
as well as the mechanical work made by the external potential on the superstar
($\Delta E$; see below).
The first criterion is satisfied if
\begin{equation}
2(E+\Delta E-\phi_{eff}(r_{t}))>\frac{L^{2}}{r_{t}^{2}}
\label{crit1_eq}
\end{equation}

The second criterion is based on the fact that the escaping star needs a
timescale comparable to its orbital period ($T$) to reach the tidal radius. This
produces a delay in the escape process. This can be crucial when eccentric orbits
are considered: in this case, $r_{t}$ changes with time and the star satisfies 
eq. \ref{crit1_eq} only in a limited time interval. If such time interval is 
short compared to its orbital period, there is only a small probability for the
star to escape. We then extracted a random number $\eta_{3}$ uniformly
distributed between 0 and 1 and calculated the
orbital period of the star to reach the tidal radius using eq.
\ref{per_eq}
. We assumed that 
the the star can escape during the time-step $\Delta t$ if
\begin{equation}
\eta_{3}<1-e^{-\frac{2\Delta t}{T}}
\label{crit2_eq}
\end{equation}

Once the first two cirteria are satisfied the star can escape from the cluster.
After this phase the star moves in the galactic potential as an independent
satellite following an epicyclic orbit along either the trailing or the leading arm of the
cluster tidal tails. During this motion, the distance of the star from the cluster 
center oscillates according to the orbital phase of the cluster. 
To definitively escape from the cluster attraction there is a typical timescale
which depends on the star's energy and on the cluster orbital parameters. 
Moreover, if the cluster follows an eccentric orbit its tidal radius grows when the
cluster leaves the perigalacticon and it can possibly exceed the distance of the
previously escaped star. In this case the star is re-captured.
To account for this effect, when a star satisfies the two above mentioned criteria its
position and velocity are calculated using eq. \ref{pos_eq} ($x_{i}=r_{t}\tilde{x}_{i}$) 
and
\begin{eqnarray}
v_{x}&=&v_{r}\tilde{x}-v_{t,a}\frac{\tilde{y}}{\sqrt{\tilde{x}^{2}+\tilde{y}^{2}}}-v_{t,b}\frac{\tilde{x}\tilde{z}}{\sqrt{\tilde{x}^{2}+\tilde{y}^{2}}}\nonumber\\
v_{y}&=&v_{r}\tilde{y}+v_{t,a}\frac{\tilde{x}}{\sqrt{\tilde{x}^{2}+\tilde{y}^{2}}}-v_{t,b}\frac{\tilde{y}\tilde{z}}{\sqrt{\tilde{x}^{2}+\tilde{y}^{2}}}\nonumber\\
v_{z}&=&v_{r}\tilde{z}+v_{t,b}\sqrt{\tilde{x}^{2}+\tilde{y}^{2}}.
\label{vel_eq}
\end{eqnarray}
where 
\begin{eqnarray*}
v_{r}&=&\sqrt{2(E+\Delta E-\phi(r_{t}))-\frac{L^{2}}{r_{t}^{2}}}\nonumber\\
v_{t,a}&=&\frac{|L|}{r_{t}} sin(2\pi\eta_{4})\nonumber\\
v_{t,b}&=&\frac{|L|}{r_{t}} cos(2\pi\eta_{4})\nonumber.
\end{eqnarray*}
and $\eta_{4}$ is a random number uniformly distributed between 0 and 1.
The above phase-space coordinates are transformed in the galactic
reference system, added to the cluster systemic coordinates and the orbit of both the star and the cluster within the 
galactic potential are
followed using a fourth-order Hermite integrator with an adaptive timestep for an entire cluster
orbital period. 
During its motion outside the cluster the
distance of the star from the cluster and the tidal radius are calculated. 
The star is removed from the simulation
if the conditions 
\begin{eqnarray}
\sqrt{(x-x_{cl})^{2}+(y-y_{cl})^{2}+(z-z_{cl})^{2}}&>&r_{t,max}\nonumber\\
\eta_{3}&<&1-e^{-\frac{\Delta t}{t_{esc}/2+T}}
\label{crit3_eq}
\end{eqnarray}
are satisfied during a cluster orbital period.
In the above equations $r_{t,max}$ is the maximum tidal radius reached by the
cluster during an orbital period, $\Delta t$ is the time-step, $T$ is the star
orbital period within the cluster and $\eta_{3}$ is the random number extracted
for the criterion in eq. \ref{crit2_eq}.   

The last modification regards the time-step adopted in the
simulation when the external field is present. In Joshi et a. (2001) the
time-step is defined to ensure small deflection angles when introducing the
perturbations to the supertars energies and angular momenta. However, another
requirement is that the Jacobi integral should not significantly vary within the
time-step. So, we adoted as time-step the minimum between the time-step defined 
by Joshi et al. (2001) and 0.01$P_{cl}$, where $P_{cl}$ is the cluster orbital
period.

\subsection{Tidal shocks}
\label{shocks_sec}

When a cluster passes through the galactic disk or close to the galactic bulge the
gravitational field of these two components exerts a compressive force which is
superposed on the cluster's own gravitational field. This process is known as
"disk/bulge shocking" (Ostriker, Spitzer \& Chevalier 1972; Aguilar, Hut \&
Ostriker 1988). The theory of tidal shocks has been studied in the past by many
authors and applied to orbit-averaged Fokker-Planck codes by Gnedin \& Ostriker 
(1997, 1999) and Allen, Moreno \& Pichardo (2006). 

This effect can be viewed as a consequence of the mechanical work made by the 
external potential during the orbit. Indeed, any variation of the potential
shape during the cluster evolution produces a work on the stars which is equal 
to the instantaneous potential variation. The variation of the internal 
potential can be due to many processes (secular dynamical evolution, 
evaporation of stars, stellar evolution, etc.) and is
taken into account using the prescriptions by Stodolkiewicz (1982).
In time-dependent external potentials (e.g.
when an eccentric orbit and/or non-spherical potentials are considered)  
also the external potential makes a work on the stars. 
This is particularly important when fast changes in the external potential
occurs, like in the case of the disk crossing and the perigalactic passages.
In this case, the binding energy per unit mass of the superstars located in the
outskirts of the cluster changes on a timescale comparable to the dynamical
time. This effect facilitates the escape of stars and can accelerate the cluster
dynamical evolution.

In analogy with 
Stodolkiewicz (1982), at each time-step we estimated the work made by the 
external potential on each
superstar as the average between the potential changes at the positions of the
superstar in two subsequent time-steps
\begin{eqnarray*}
\Delta E&=&[\phi_{ext}(r'(t),t)+\phi_{ext}(r'(t-\Delta t),t)-\nonumber\\
& & \phi_{ext}(r'(t),t-\Delta t)-\phi_{ext}(r'(t-\Delta t),t-\Delta t)]/2\\
\end{eqnarray*}
Such a correction is updated during the orbit and temporary added to the 
energy of the superstar only to verify the escape criteria defined above. 
This correction cannot indeed be permanently added the superstars energies since 
these are distributed across the cluster adopting the internal potential only 
(i.e. assuming the cluster as isolated) while the external potential is taken 
into account only in a subsequent step when the escape criteria are verified.
 
\section{Comparison with N-body simulations}
\label{res_sec}

\begin{table*}
 \centering
 \begin{minipage}{140mm}
  \caption{Summary of the performed simulations.}
  \begin{tabular}{@{}lcccccccccr@{}}
  \hline
  simulation & external potential & N & $e$ & $r_{cl,ap}$ & $z_{cl,max}$ & cluster profile &
  $W_{0}$  & $r_{c}$  & $\log{M_{cl}/M_{\odot}}$ & $\langle\mu\rangle$\\
             &                    &   &     & kpc         & kpc          &                 & 
	   &  pc      & &\\
 \hline
 P-iso        & isolated    & 8192    & -- & -- & -- & Plummer & 1\footnote{For the
 P-iso simulation the Plummer characteristic radius is indicated.} & -- & 4 & --\\       
 K-pm-e0-21K     & point-mass & 21000 & 0.0   & 4 & 0  & King    & 5 & 2  & 4 & -0.00007\\	   
 K-pm-e014-21K   & point-mass & 21000 & 0.143 & 4 & 0  & King    & 5 & 2  & 4 & -0.00011\\	   
 K-pm-e033-21K   & point-mass & 21000 & 0.333 & 4 & 0  & King    & 5 & 2  & 4 & -0.00022\\	
 K-pm-e060-21K   & point-mass & 21000 & 0.6   & 4 & 0  & King    & 5 & 2  & 4 & -0.02428\\	
 K-j95-e0-21K    & J95        & 21000 & 0.0   & 10 & 6 & King    & 5 & 2  & 4 & -0.00006\\	   
 K-j95-e033-21K  & J95        & 21000 & 0.0   & 10 & 6 & King    & 5 & 2  & 4 & -0.00010\\	   
 K-j95-e033z-21K & J95        & 21000 & 0.0   & 10 & 8 & King    & 5 & 2  & 4 & -0.00009\\	   
 K-pm-e0-r2-21K  & point-mass & 21000 & 0.0   & 2 & 0  & King    & 5 & 2  & 4 & -0.00040\\	   
 K-pm-e0-r3-21K  & point-mass & 21000 & 0.0   & 3 & 0  & King    & 5 & 2  & 4 & -0.00013\\	   
 K-pm-e0-42K     & point-mass & 42000 & 0.0   & 4 & 0  & King    & 5 & 2  & 4 & -0.00004\\	   
 K-pm-e014-42K   & point-mass & 42000 & 0.143 & 4 & 0  & King    & 5 & 2  & 4 & -0.00006\\	   
 K-pm-e033-42K   & point-mass & 42000 & 0.333 & 4 & 0  & King    & 5 & 2  & 4 & -0.00015\\	
 K-pm-e060-42K   & point-mass & 42000 & 0.6   & 4 & 0  & King    & 5 & 2  & 4 & -0.23070\\	
 K-j95-e0-42K    & J95        & 42000 & 0.0   & 10 & 6 & King    & 5 & 2  & 4 & -0.00004\\	   
 K-j95-e033-42K  & J95        & 42000 & 0.0   & 10 & 6 & King    & 5 & 2  & 4 & -0.00006\\	   
 K-j95-e033z-42K & J95        & 42000 & 0.0   & 10 & 8 & King    & 5 & 2  & 4 & -0.00006\\	   
\hline
\end{tabular}
\end{minipage}
\end{table*}

\begin{figure}
 \includegraphics[width=8.7cm]{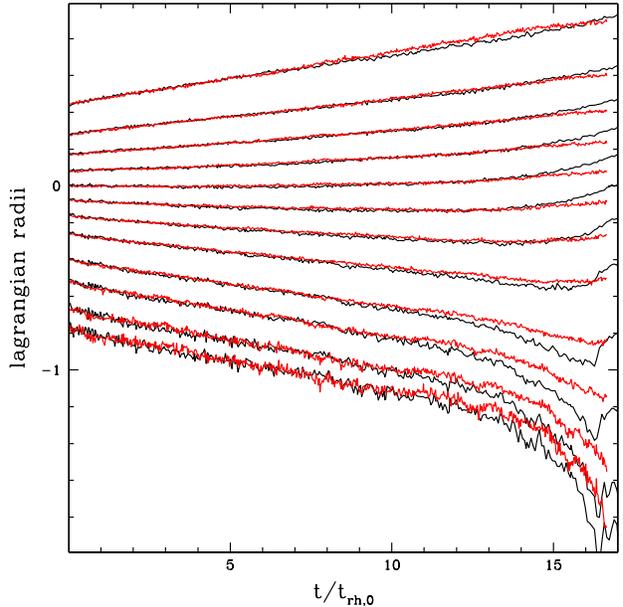}
 \caption{Comparison between the evolution of the Lagrangian radii predicted by the Monte Carlo code presented here (red
 lines; grey in the printed version of the paper) and the N-body simulations by
 Baumgardt et al. (2002; black lines) for the
 isolated Plummer (1911) model. The radii containing 1, 2, 5, 10, 20, 30, 40, 50, 60,
 70, 80, 90\% of the cluster mass are shown.}
\label{plum}
\end{figure}

\begin{figure}
 \includegraphics[width=8.7cm]{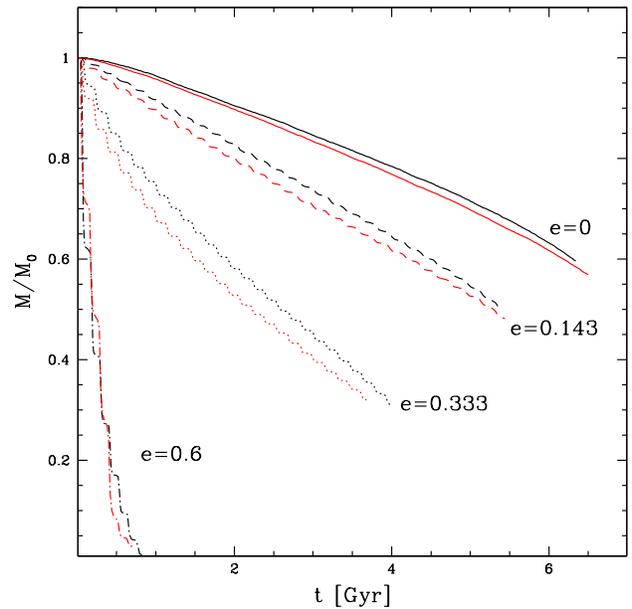}
 \caption{Comparison between the mass evolution predicted by the Monte Carlo 
 code presented here (red lines; grey in the printed version of the paper) and 
 the N-body simulations (black lines) for the
 King model orbiting around a point-mass galaxy with 21000 particles. Simulations with eccentricity
 $e$=0, 0.14, 0.33 and 0.6 are drawn with solid, dashed, dotted and dot-dashed
 lines, respectively.}
\label{pm_mass}
\end{figure}

\begin{figure*}
 \includegraphics[width=12cm]{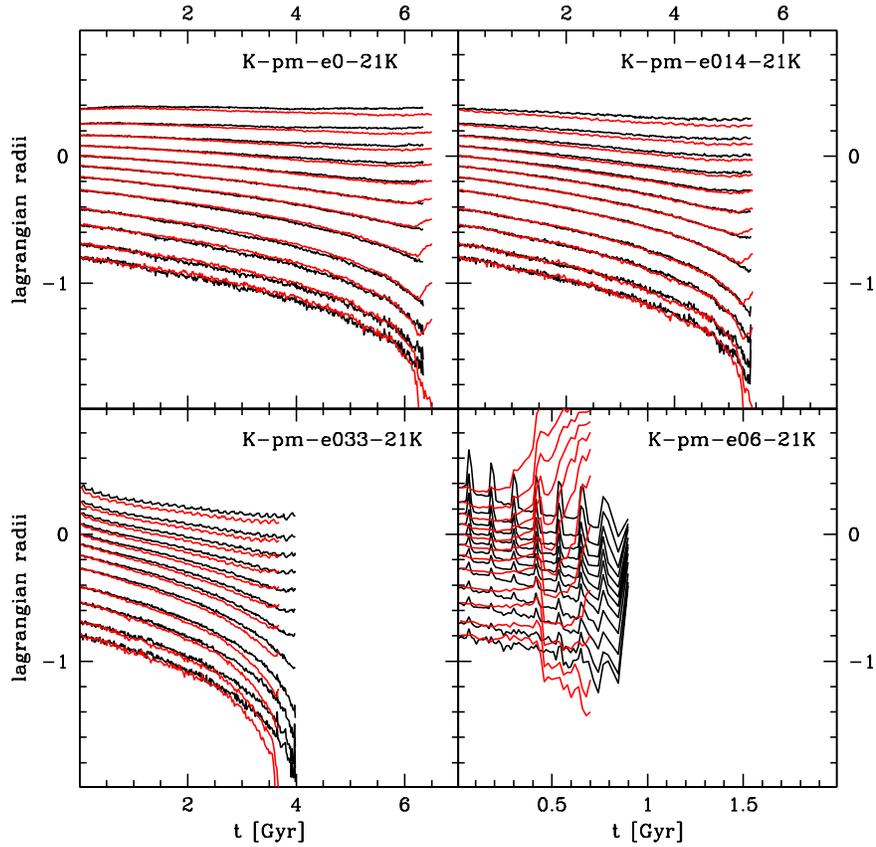}
 \caption{Comparison between the evolution of the Lagrangian radii predicted by 
 the Monte Carlo code presented here (red lines; grey in the printed version of the paper) and 
 the N-body simulations (black lines) for the
 King model orbiting around a point-mass galaxy with 21000 particles and
 different eccentricities.}
\label{pm_lag}
\end{figure*}

\begin{figure}
 \includegraphics[width=8.7cm]{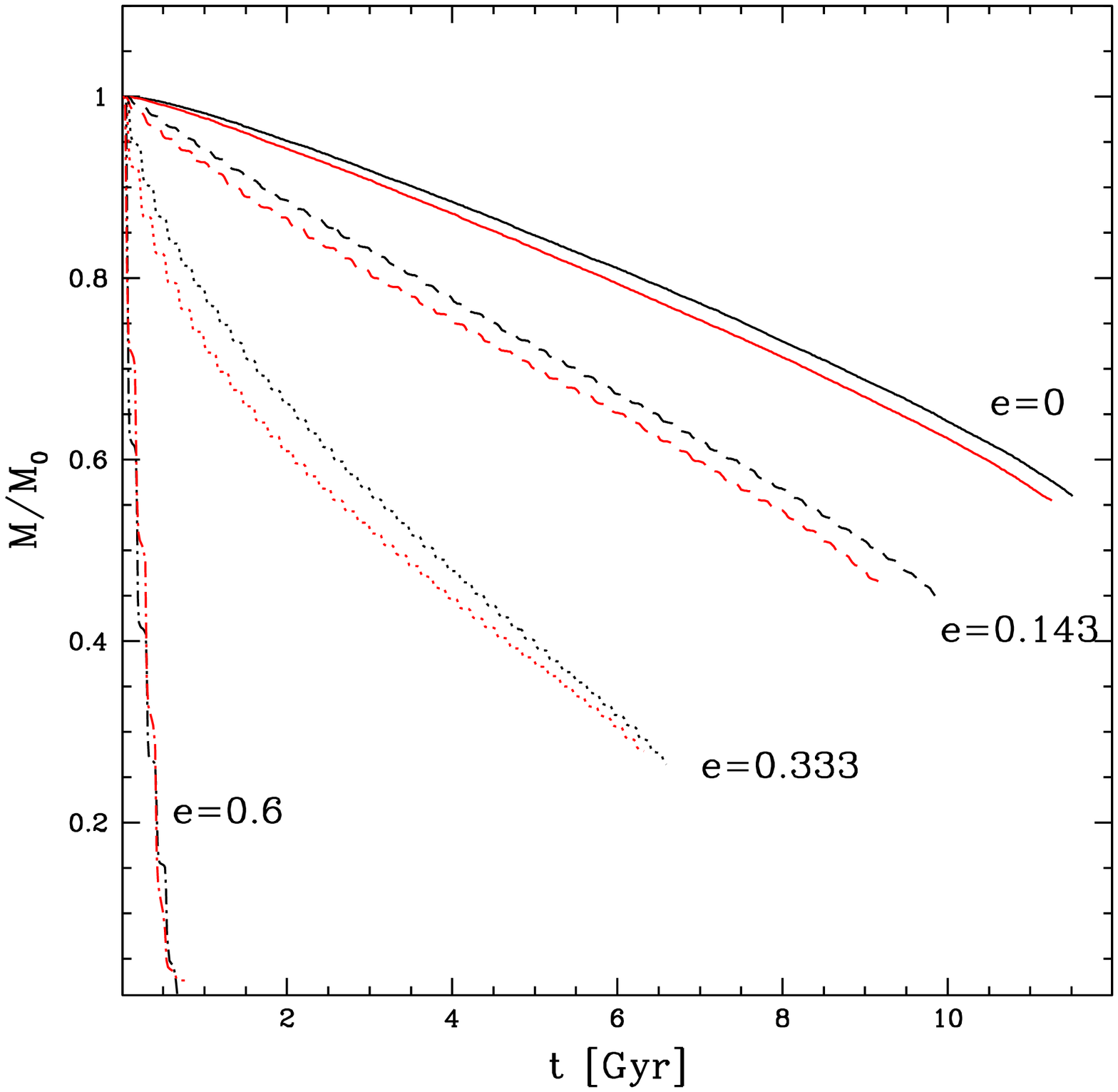}
 \caption{Same as Fig. \ref{pm_mass} but for the set of simulations with 42000 particles.}
\label{pm_mass42}
\end{figure}

\begin{figure*}
 \includegraphics[width=12cm]{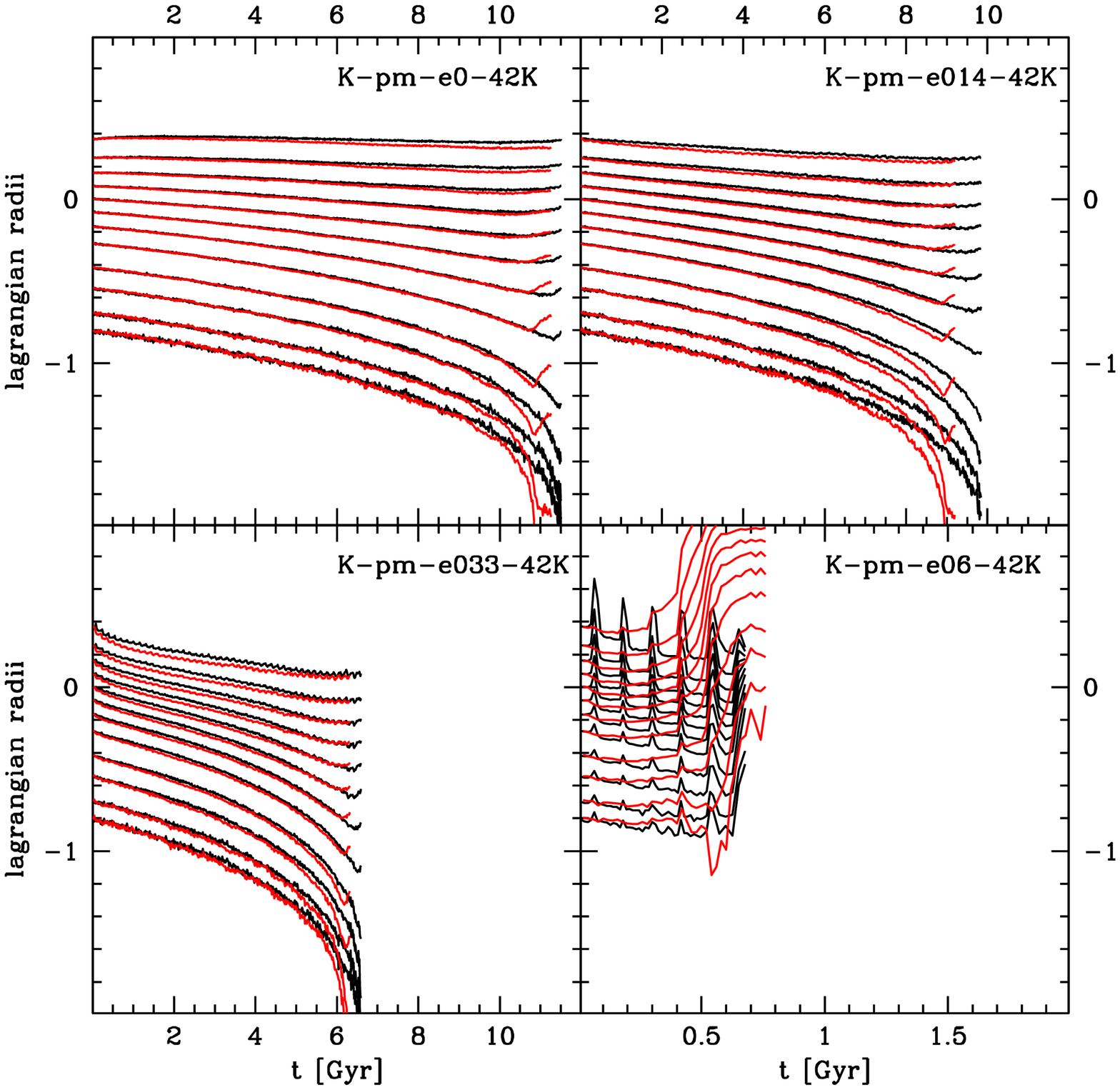}
 \caption{Same as Fig. \ref{pm_lag} but for the set of simulations with 42000 particles.}
\label{pm_lag42}
\end{figure*}

\begin{figure}
 \includegraphics[width=8.7cm]{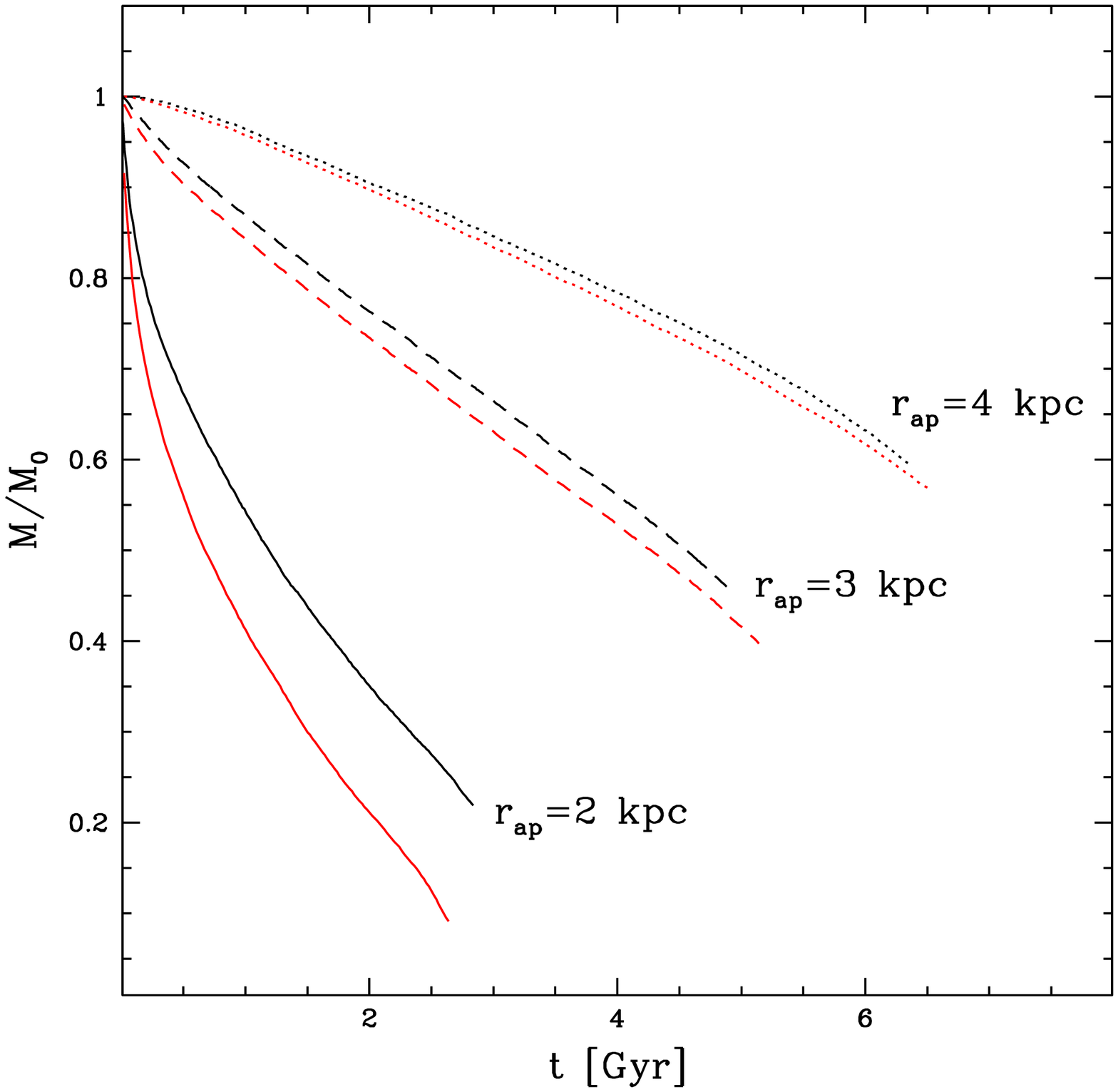}
 \caption{Comparison between the mass evolution predicted by the Monte Carlo 
 code presented here (red lines; grey in the printed version of the paper) and 
 the N-body simulations (black lines) for the
 King model orbiting around a point-mass galaxy with 21000 particles.
 Simulations with distances from the point-mass galaxy 
 $r_{ap}$=2, 3, and 4 kpc are drawn with solid, dashed, and dotted
 lines, respectively.}
\label{pm_testmass}
\end{figure}

We tested the prediction of our code with a set of collisional N-body simulations performed
with {\rm NBODY6} (Aarseth 1999). All simulations (both Monte Carlo and N-body) 
were run with N=21000 and N=42000 single-mass particles as a compromise to ensure a large
number statistics and to limit the computational cost of N-body simulations.
The standard $\eta=0.02$ parameter has been used to control the time step and 
set an energy error tolerance of $2.0\times10^{-4}$. With these choices we got a
relative error in energy smaller than $\Delta E/E\sim4\times 10^{-4}$ 
at the end of every simulation run. For each simulation the cluster mass
contained within the apocentric tidal radius and the Lagrangian radii have been calculated and compared.
Lagrangian radii have been calculated from the cumulative distribution of distances 
from the cluster center of the particles. 
Here we considered the radii containing 1\%, 2\%, 5\%, 10-90\% of the cluster mass 
within one apocentric tidal radius.

We considered three different cases: {\it i)} an isolated cluster with a Plummer
(1911) profile; {\it ii)} a cluster with a mass $M=10^{4}~M_{\odot}$, a King (1966) profile 
with $W_{0}=5$ and $r_{c}=2~pc$ orbiting in a galactic potential
generated by a point-mass of $M_{g}=10^{10}~M_{\odot}$ with an apogalacticon at
$r_{cl,ap}=4~kpc$, and {\it iii)} the same
cluster of (ii) orbiting in the bulge+disk+halo galactic potential defined by
J95. In cases {\it (ii)} and {\it (iii)} a set of simulations
with different eccentricities and orbits have been considered. The whole set of
simulations is summarized in Table 1. 

In Fig. \ref{plum} the evolution of the Lagrangian radii of the isolated Plummer
(1911) model as a function of the initial half-mass relaxation time is compared with 
that predicted by the N-body simulation with N=8192 particles by Baumgardt, Hut
\& Heggie (2002). The agreement is excellent with only a small discrepancy near
the core collapse for the innermost radii. 
The core collapse occurs after $\sim16.5~t_{rh,0}$ when the cluster have lost
$\sim$3\% of its stars. Both quantities are in good
agreement with the results of Baumgardt et al. (2002). The excellent agreement
with the N-body simulation indicates that the Monte Carlo code well reproduces
the relaxation process until the core collapse.

\begin{figure*}
 \includegraphics[width=15cm]{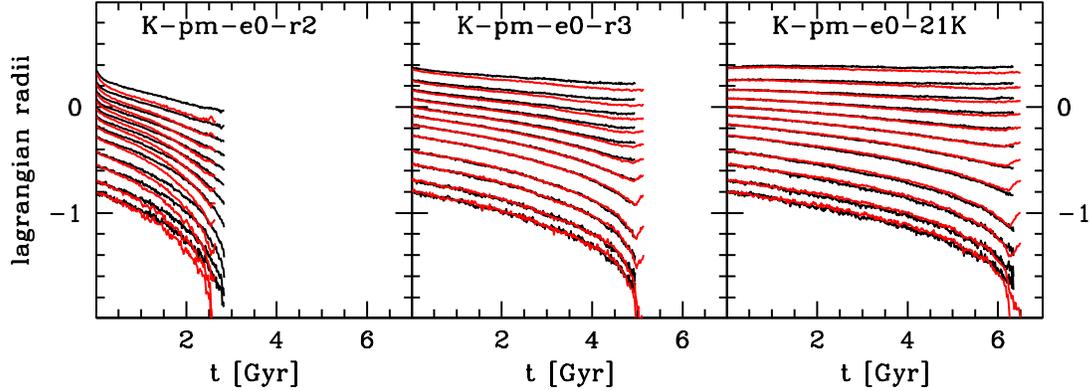}
 \caption{Comparison between the evolution of the Lagrangian radii predicted by 
 the Monte Carlo code presented here (red lines; grey in the printed version of the paper) and 
 the N-body simulations (black lines) for the
 King model orbiting around a point-mass galaxy with 21000 particles and
 different distances from the point-mass galaxy.}
\label{pm_testlag}
\end{figure*}

Another set of simulations have been performed considering an external tidal 
field generated by a point-mass galaxy. In such a potential stars feel the tidal cut
but are not subject to disk shocks. These simulations are therefore suited to
test the escape from the cluster boundary both in case of circular and eccentric
orbits. Simulations have been run until core collapse. The subsequent
evolution is largely influenced by the presence of binaries which form
during the maximum density phase. As our Monte Carlo code does still not account
for this process it cannot reproduce properly such an evolutionary stage.
The evolution of the cluster mass and of the Lagrangian radii for this 
set of simulations are compared with the results of N-body simulations in 
Fig. \ref{pm_mass} and \ref{pm_lag}, respectively. It is apparent that in all the simulations
with eccentricity $e<0.5$ the Monte Carlo tends to systematically 
overpredict the mass-loss
rate with respect to N-body simulations, although such a discrepancy is always 
within 5\%. The evolution of the Lagrangian radii is also well reproduced
during the entire cluster evolution.
A different situation is for the simulation of the most eccentric ($e=0.6$) 
orbit. In this case, while the evolution of the bound mass is well reproduced by
our Monte Carlo code, the
Lagrangian radii are strikingly different. In particular, while
the N-body simulation predict an overall contraction of the cluster, the
Monte Carlo code predicts a quick expansion followed by a quick collapse of the
core. A possible reason for such a
discrepancy is that in this last orbit the tidal radius penetrates into the
cluster at pericenter leaving a significant fraction of cluster stars free to
escape. When this occurs, the criteria defined in Sect. \ref{esc_sec} are not
adequate anymore and an unrealistically large fraction of stars evaporate from
the cluster in a short amount of time. The loss of potential energy is 
larger than that in kinetic energy and the cluster expands, furthermore
increasing the escape efficiency. 
As the number of stars in the cluster becomes
smaller the relaxation process speeds up and the core quickly collapse. 
Summarizing, it appears that the treatment of the external tidal field described
in Sect. \ref{esc_sec} is
effective when the escape rate of stars during a cluster orbital period is 
smaller than a critical value. 
In Fig. \ref{pm_mass42} and \ref{pm_lag42} the comparison between the bound
mass and Lagrangian radii for the same 
set of simulations with 42000 particles are shown, respectively.
Also in this case, the agreement is good in all the simulations with moderate
eccentricity ($e<0.5$) both in terms of the mass and of the Lagrangian
radii evolution. Again, for the most eccentric simulation, while the bound mass 
evolution is fairly well reproduced, the Lagrangian radii of the Monte Carlo
simulation show the same unproper behaviour already noticed in the simulations
with 21000 particles.

\begin{figure*}
 \includegraphics[width=12cm]{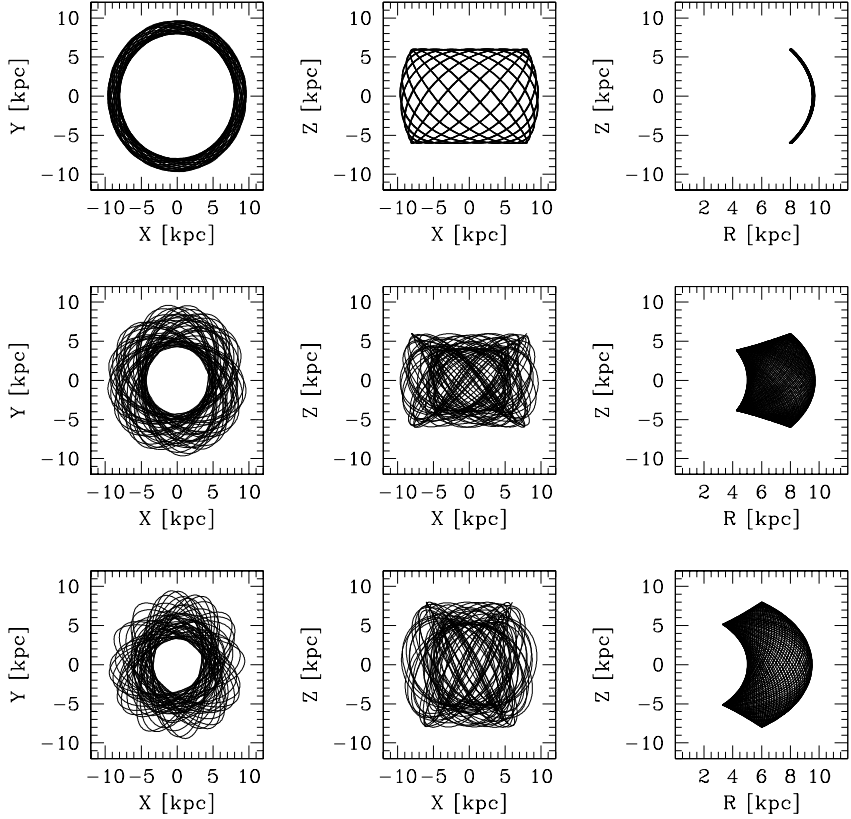}
 \caption{Orbits followed by the cluster during 12 Gyr for the simulations K-j95-e0 (top panels), K-j95-e033
 (middle panels) and K-j95-e033z (bottom panels). The left, central and right
 panels show the orbit in the X-Y, X-Z and R-Z planes, respectively.}
\label{orb}
\end{figure*}

To properly estimate the range of validity of our code, we performed another 
set of simulations where a cluster with 21000 particles and the same
structural characteristics of the previous simulations
has been launched on circular orbits at different distances from the point-mass 
galaxy ($r_{ap}=$2,3 and 4 kpc). The results of such an experiment are shown
in Fig.s \ref{pm_testmass} and \ref{pm_testlag}. As expected, Monte Carlo 
simulations related to clusters moving at large galactocentric distances show a 
good agreement with N-body ones. On the other hand, as the tidal field becomes
stronger the mass-loss rate predicted by the Monte Carlo code exceeds that that 
of the N-body simulations. As a consequence, the cluster internal dynamical evolution 
accelerates and the core collapse is anticipated. According to the escape
criteria defined in Sect. \ref{esc_sec}, the timescale at which a star escapes
from the cluster is of the order of a few dynamical times. While this
timescale depends on the energy and angular momentum of each star and on the
cluster potential, it is expected to scale with the global quantity
$t_{dyn,r_{h}}\propto\sqrt{r_{h}^{3}/G M}$. It is therefore useful to introduce
the mass-loss rate per half-mass dynamical time
$$\mu=\frac{\dot{M}}{M}\sqrt{\frac{r_{h}^{3}}{G M}}$$
The above parameter has been calculated during the cluster evolution and
averaged from the beginning of the simulation to the core collapse (see
 Table 1). 
To define a criterion of validity of our simulations we correlated the values of 
$\langle\mu\rangle$ with the discrepancy between bound mass fraction predicted
by Monte Carlo and N-body simulations
measured at core-collapse in this set of simulations. We obtain
$\langle\mu\rangle$=-0.00007, -0.00013 and -0.00040 and $\Delta M/M_{0}$=1.2\%,
3.0\% and 15.9\% for the simulations at 4, 3 and 2 kpc, respectively. On the
basis of the above comparison and defining a resonable agreement at $\Delta
M/M_{0}<$5\%, we adopt
a value of $\langle\mu\rangle>$-0.0002 as a conservative limit of validity of 
our code.

\begin{figure*}
 \includegraphics[width=12cm]{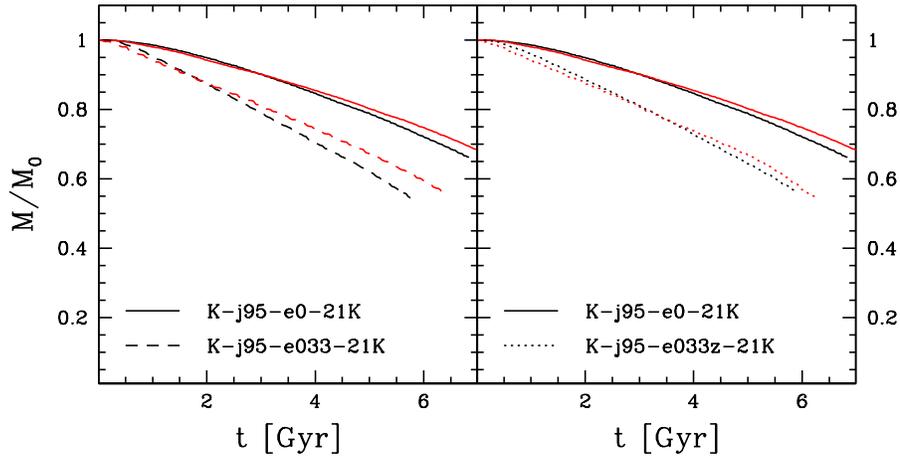}
 \caption{Same as Fig. \ref{pm_mass} for the
 King model orbiting within the J95 galactic potential. Simulations K-j95-e0-21K
 (both panels), K-j95-e033-21K (left panel) and K-j95-e033z-21K (right panel) are drawn with solid, dashed, dotted and dot-dashed
 lines, respectively.}
\label{compl_mass}
\end{figure*}

\begin{figure*}
 \includegraphics[width=12cm]{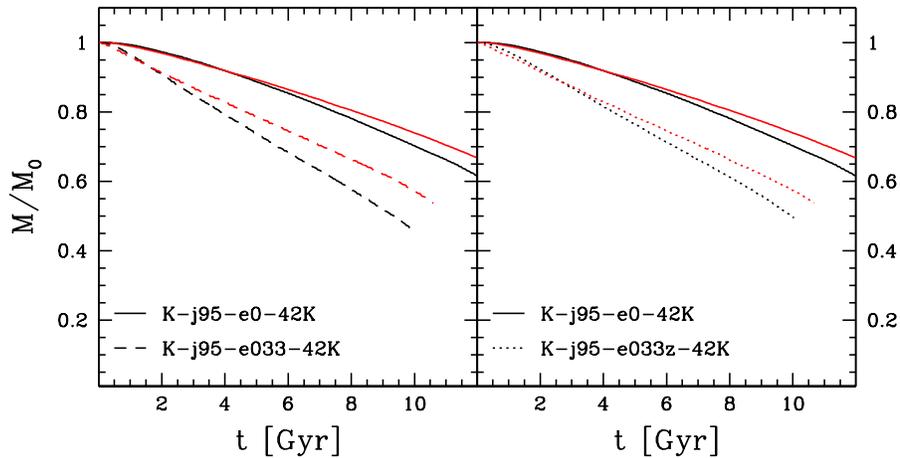}
 \caption{Same as Fig. \ref{compl_mass} but for the set of simulations with 42000 particles.}
\label{compl_mass42}
\end{figure*}

\begin{figure*}
 \includegraphics[width=15cm]{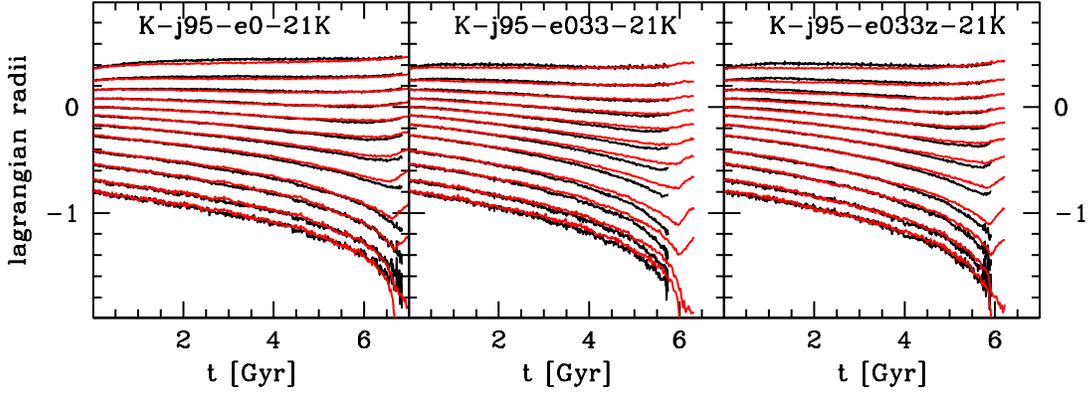}
 \caption{Same as Fig. \ref{pm_lag} for the
 King model orbiting within the J95 galactic potential.}
\label{compl_lag}
\end{figure*}

\begin{figure*}
 \includegraphics[width=15cm]{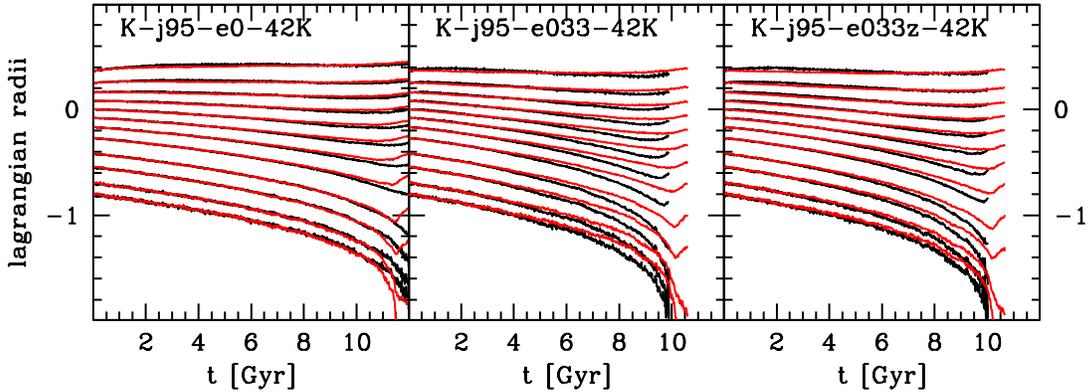}
 \caption{Same as Fig. \ref{compl_lag} but for the set of simulations with 42000 particles.}
\label{compl_lag42}
\end{figure*}

The last set of simulations considers a cluster immersed in the three-components
bulge+disk+halo external potential defined by J95. We considered a cluster lying
at an initial distance of 10 kpc from the galactic center with two different
eccentricities and two different heights above the galactic plane. In each case
the intensities of the tidal shocks are different since the cluster cross the disk
and approaches its pericenter at different distances from the galactic center 
with different velocities. The three considered orbits in the X-Y, X-Z and R-Z
planes are shown in Fig. \ref{orb}. The mass evolution of the three simulations
are compared with those predicted by the
associated N-body simulations in Fig.s \ref{compl_mass} and
\ref{compl_mass42} for the set with 21000 and 42000 particles, respectively. 
It can be noted that in all the simulations with 21000 particles
the agreement is good (within 5\%) during the entire evolution.
On the other hand, in simulations with 42000 particles a tendency of the
Monte Carlo code to underestimate the cluster mass-loss rate is noticeable. This
is particularly apparent in the simulations with eccentric orbits where the
difference with respect to the prediction of the N-body simulation reach
$\sim$10\% at the core-collapse. Such a discrepancy is in the opposite sense of what observed in
simulations run within a point-mass potential, where the mass-loss rates were
slightly overpredicted.
Note that in the three simulations the average mass-loss rate 
is $\langle\mu\rangle>-0.0001$ i.e. below the critical limit where significant
differences have been noticed in the simulations within a point-mass external
potential. 
It is also interesting to note that the mass evolution of the two simulations
with different heights above the galactic plane (K-j95-e033 and K-j95-e033z) 
yield to a quite similar residual mass after 12 Gyr. This is not surprising since 
at the moment of the disk crossing, although the disk shocks are
more intense in simulation K-j95-e033z because of the largest velocity of the
cluster, the disk density is in both cases relatively small. Also, the
pericentric distance of both orbits is 5 kpc, significantly larger with respect
to the bulge half-mass radius ($\sim$1.69 kpc). So, in the above cases, both
disk and bulge shocks have only a little impact on the cluster structural and
dynamical evolution.
In Fig.s \ref{compl_lag} and \ref{compl_lag42} the evolution of the Lagrangian radii of the three
simulations are compared with the predictions of the corresponding N-body
simulations for the sets with 21000 and 42000 particles, respectively. Again the agreement is good during the entire cluster evolution.

\section{Summary}

In this paper we presented a new implementation of the Monte Carlo method to
simulate the evolution of star clusters including for the first time the effect 
of a realistic tidal field and the possibility to consider eccentric orbits.
The effect of the external field has been taken into account considering both
the process of evaporation through the cluster boundaries and the effect of tidal
shocks. The adopted algorithm is based on the theory of evaporation taking into 
account for the random direction of the escape, the time delay due to the star 
orbital period and the occurrence of the re-capture process. 

The comparison with direct N-body simulations indicates an excellent
agreement
in an isolated cluster and a good agreement (within 5-10\%) in clusters within a tidal field of moderate
intensity, in terms of the evolution of both the mass and the Lagrangian radii. 
This indicates that both the process of two-body relaxation and the
evaporation of stars are accurate enough.
On the other hand, a
significant discrepancy is apparent when an extremely intense tidal stress is
present ($\dot{M}/M/t_{dyn,r_{h}}>-0.0002$). The reason of the disagreement relies on
the failure of some (maybe all) of the criteria adopted to take into account 
the escape of stars when these apply to a significant fraction of cluster stars.
It is worth noting that such an intense tidal field should be felt by
only a small fraction of globular clusters (only 2 out the 53 globular clusters
considered by Allen et al. 2006, 2008). So this code can provide an efficient
tool to study the evolution of present-day globular clusters.
On the other hand, the code developed here could fail to reproduce the initial
stages of cluster evolution: in this case, the fast potential 
changes due to stellar evolution driven mass loss likely lead to a situation of
Roche-lobe overfilling where the critical mass loss rate is easily reached.
Consider that one of the basic
assumptions of the Monte Carlo method is spherical symmetry which conflicts with
the presence of the external field. In this situation star orbits are not
expected to be planar and it is not guaranteed that two superstars with
contiguous ranking in distance from the cluster center are neighbors. 
So, any treatment of the external field is expected to fail when somehow strong
tides are considered. 

The performance of the code are also very good: the simulation K-pm-e0-21K presented here 
with 21000 particles takes $\sim$45 minutes ("clock on the wall" time) on a single 
{five-years old ASUS machine equipped with a single Intel Core T5800@2GHz processor, 
while the corresponding N-body simulation takes
$\sim$30 hours with a cluster node equipped with an Intel Xeon E5645 CPU (12
cores) and a NVIDIA Tesla M2090 (512 CUDA cores)}.
In spite of the relatively small considered number of particles, the above 
performance is good considering that the code include several cycles
which can be in principle easily parallelized thus reducing the computational
cost of simulations. Assuming a scaling of the computation time with the number 
of particles as $\propto N~log_{2}N$, it is possible to run simulations with
$N\sim10^{6}$ with computation time of few days.
 
The code already include other features to properly simulate the cluster
evolution like the inclusion of a mass spectrum, the mass-loss driven by stellar evolution and the direct
integration of three- and four-body interactions following the prescriptions by
Joshi et al. (2001) and Fregeau et al. (2007). However, the predictions of the 
code with these implemented features have not been tested and are not presented
here. A forthcoming paper will introduce these features in the next future.

\section*{Acknowledgments}

AS acknowledges the PRIN INAF 2011 "Multiple populations in globular
clusters: their role in the Galaxy assembly" (PI E. Carretta). AMB whish to
thank the Lady Davis Foundation. We thank the anonymous referee for his/her
helpful comments and suggestions. We warmly thank
Holger Baumgardt to have provided his N-body simulations and Zeinab Khorrami, Carlo Nipoti, Luca
Ciotti and Enrico Vesperini for useful discussions.

\appendix
\onecolumn

\section[]{Tidal radius and effective potential within an external tidal field}

As discussed in Sect. \ref{esc_sec}, when the cluster is immersed in an external
tidal field its stars feel an effective potential due to the combination of the
internal cluster potential ($\phi_{cl}$), the external field potential
($\phi_{ext}$) and a term linked to the cluster angular motion. Such an effective 
potential does not have a spherical symmetry so that the distances of its local 
maxima (tidal radii) are direction dependent. In the next sections the tidal
radius and the effective potential as a function of the direction of escape are
derived both for the case of an external field generated by a point-mass and for
the complex bulge+disk+halo potential by J95.

\subsection{Point-mass galaxy}
\label{pm_sec}

The external potential generated by a point-mass galaxy is
$$\phi_{ext}=-\frac{G M_{g}}{r}$$
where $M_{g}$ is the mass of the galaxy and $r$ the distance of a test
particle from the point-mass. Consider a star at $r$ orbiting around a cluster
located at $r_{cl}$. In the Cartesian reference frame centered on the cluster with the
x-axis pointed toward the point-mass galaxy, the y-axis in the direction of the
rotation and the z-axis in the direction of the angular momentum, be ${\bf
r'}=(x,y,z)$ the position vector of the star.
The distance of the star from the galaxy will be 
$$r=\sqrt{r_{cl}^{2}+|{\bf r'}|^{2}-2 x r_{cl}}$$

The tidal radius is defined as the point where the
projection of the acceleration felt by the star (eq. \ref{acc_eq}) on ${\bf
r'}$ is zero. The acceleration is composed by three terms: one associated to the
cluster potential, one associated to the external field and another associated
to the cluster angular motion.

At the tidal radius the potential of the cluster is that of a point-mass, and 
the internal acceleration has a projection
on ${\bf r'}$ 
\begin{equation}
\frac{{\bf r'}}{|{\bf r'}|} \cdot {\bf \nabla\phi_{cl}}=\frac{G M_{cl}}{|{\bf r'}|^{2}}
\label{pmc_eq}
\end{equation}.

The term associated to the external field is
\begin{eqnarray*}
\frac{{\bf r'}}{|{\bf r'}|} \cdot ({\bf r'} \cdot \nabla) \nabla \phi_{ext}&=&
\frac{G M_{g}}{|{\bf r'}|} \sum_{i=1}^{3} \frac{x_{i}^{2}}{|{\bf r}|^{2}} \left[\frac{\delta^{2}
r}{\delta x_{i}^{2}}-\frac{2}{|{\bf r}|}\left(\frac{\delta r}{\delta
x_{i}}\right)^{2}\right]\nonumber\\
 &=& \frac{G M_{g}}{|{\bf r'}| |{\bf r}|^{5}}[x^{2}\left(-2 r_{cl}^{2}+|{\bf
r'}|^{2}+4 x r_{cl}-3 x^{2}\right)+y^{2}\left(r_{cl}^{2}+|{\bf
r'}|^{2}-2 x r_{cl}-3 y^{2}\right)+z^{2}\left(r_{cl}^{2}+|{\bf
r'}|^{2}-2 x r_{cl}-3 z^{2}\right)]\nonumber
\end{eqnarray*}
In the limit $|{\bf r'}|<<r_{cl}$ the above term can be approximated to
\begin{equation}
\frac{{\bf r'}}{|{\bf r'}|} \cdot ({\bf r'} \cdot \nabla) \nabla \phi_{ext}=-\frac{G
M_{g}}{|{\bf r'}| r_{cl}^{3}}(2x^{2}-y^{2}-z^{2})
\label{pmf_eq}
\end{equation}.

The acceleration due to the angular motion of the cluster is
\begin{equation}
{\bf a_{\Omega}}=2 {\bf \Omega\times{v'}}+{\bf
\Omega\times (\Omega\times r')}+{\bf \frac{d \Omega}{dt}\times r'}
\label{pmat_eq}
\end{equation}
Where $\Omega$ is the angular speed of the cluster which, in an elliptic orbit
with eccentricity $e$ and apogalactic distance $r_{cl,ap}$ is
$$\Omega=\sqrt{\frac{G M_{g} r_{cl,ap} (1-e)}{r_{cl}^{4}}}$$
The third term of equation \ref{pmat_eq} has null projection 
on ${\bf r'}$ since it is orthogonal to ${\bf r'}$ by
definition. In the cluster reference system the remaining terms (corresponding to
the Coriolis and the centrifugal+tidal acceleration) have projection on ${\bf r'}$
\begin{eqnarray}
\frac{{\bf r'}}{|{\bf r'}|} \cdot {\bf a_{\Omega}}&=&-\frac{2\Omega}{|{\bf r'}|}
(x~v_{y}-y~v_{x})-\frac{\Omega^{2}}{|{\bf r'}|}
(x^{2}+y^{2})\nonumber\\
&=&-\frac{G M_{g} r_{cl,ap} (1-e)}{|{\bf r'}|
r_{cl}^{4}}\left(x^{2}+y^{2}+2\frac{x~v_{y}-y~v_{x}}{\Omega}\right)
\label{pma_eq}
\end{eqnarray}.

The total acceleration in the direction of ${\bf r'}$ at $r_{t}$ will be given 
by the sum of eq.s \ref{pmc_eq}, \ref{pmf_eq} and \ref{pma_eq}
\begin{equation}
{\bf \nabla \phi_{eff}}=\frac{G M_{cl}}{|{\bf r'}|^{2}}-\frac{G M_{g}
|{\bf r'}| \beta}{r_{cl}^{3}}-\frac{2 \Omega L sin(2\pi\eta_{4})}{|{\bf
r'}|}\sqrt{\tilde{x}^{2}+\tilde{y}^{2}}
\label{pmfin_eq}
\end{equation}
where
\begin{eqnarray}
\xi&=&(1-e)\frac{r_{cl,ap}}{r_{cl}}\nonumber\\
\beta&=&(2+\xi)\tilde{x}^{2}+(\xi-1)\tilde{y}^{2}-\tilde{z}^{2}\nonumber\\
\tilde{x}_{i}&=&\frac{x_{i}}{|{\bf r'}|}.
\label{beta_eq}
\end{eqnarray}
$L$ is the angular momentum of the superstar and $\eta_{4}$ is a random number
uniformly distributed between 0 and 1 (see eq. \ref{vel_eq}).
It is interesting to note that the Coriolis acceleration is directed toward the
direction of escape when the star is on a prograde orbit, while it is directed
toward the cluster when retrograde orbits are considered. This is the reason why
stars on prograde orbits escape more easily from the cluster (see also H{\'e}non
1969; Read et al. 2006). 

The tidal radius can be found by equating eq. \ref{pmfin_eq} to zero and
assuming $|{\bf r'}|=r_{t}$.
So
$$r_{t}=\left(-\frac{q}{2}+\sqrt{\frac{q^{2}}{4}+\frac{p^{3}}{27}}\right)^{\frac{1}{3}}+
\left(-\frac{q}{2}-\sqrt{\frac{q^{2}}{4}+\frac{p^{3}}{27}}\right)^{\frac{1}{3}}$$
with
\begin{eqnarray*}
p&=&\frac{2 \Omega L sin(2\pi\eta_{4}) r_{cl}^{3}}{G M_{g} \beta}\sqrt{\tilde{x}^{2}+\tilde{y}^{2}}\nonumber\\
q&=&-\frac{M_{cl} r_{cl}^{3}}{M_{g} \beta}.
\end{eqnarray*}

Note that over a large number of random extractions of $\eta_{4}$ the term 
$L~sin(2\pi\eta_{4})$ has null mean.
In this case the tidal radius will simply be
$$r_{t}=r_{cl}\left(\frac{M_{cl}}{\beta M_{g}}\right)^{\frac{1}{3}}$$
At pericenter and in correspondence to the Lagrangian
point $(\tilde{x},\tilde{y},\tilde{z})=(1,0,0)$ will be $\beta=3+e$, and the above
equation reduces to that reported by King (1962).

The effective potential can be found as the line integral of eq.
\ref{pmfin_eq}.
$$\phi_{eff}(|{\bf r'}|)=\phi_{cl}(|{\bf r'}|)-\frac{G M_{g} \beta |{\bf r'}|^{2}}{2
r_{cl}^{3}}
$$
In $r_{t}$ the above quantity is
$$\phi_{eff}(r_{t})=-\frac{3}{2} \frac{G (M_{cl}^{2} M_{g}
\beta)^{\frac{1}{3}}}{r_{cl}}$$

\subsection{Multi-component galactic potential}

We consider the three components galactic model by J95 which consists of the
superposition of an Hernquist (1990) bulge, a Miyamoto \& Nagai (1975) disk and 
a logarithmic halo
\begin{eqnarray*}
\phi_{b}&=&-\frac{G M_{b}}{r+c}\nonumber\\
\phi_{d}&=&-\frac{G M_{d}}{\sqrt{R^{2}+(a+\sqrt{b^{2}+Z^{2}})^{2}}}\nonumber\\
\phi_{h}&=&v_{0}^{2}~ln\left(1+\frac{r^{2}}{d^{2}}\right)\nonumber\\
\end{eqnarray*}
with $M_{b}=3.4\cdot10^{10}~M_{\odot}$, $c=0.7~kpc$, $M_{d}=10^{11}~M_{\odot}$,
$a=6.5~kpc$, $b=0.25~kpc$, $v_{0}=128~km/s$ and $d=12~kpc$.
Here we adopted a convenient cylindrical reference system $(R,Z,\varphi)$,
$r=\sqrt{R^{2}+Z^{2}}$ is the distance from the galactic center and $\theta$ is
the angle between the Z-axis and the position vector.
In this reference system the cluster will be in $(R_{cl},Z_{cl},\varphi_{cl})$.
Define an alternative Cartesian rotating reference frame centered in the
cluster center with the x-axis pointed toward the galactic center, the y-axis
parallel to the galactic plane and pointed toward the direction of the cluster
rotation and the z-axis perpendicular to the other axes. Consider a star in ${\bf r'}=(x,y,z)$ moving 
around a cluster immersed in the above external potential. The coordinate
transformation to the first reference system are
\begin{eqnarray*}
R&=&\sqrt{(R_{cl}-x~sin\theta_{cl}-z~cos\theta_{cl})^{2}+y^{2}}\nonumber\\
Z&=&Z_{cl}-x~cos\theta_{cl}+z~sin\theta_{cl}\nonumber\\
\varphi&=&tan^{-1}\left(\frac{R_{cl}~sin\varphi_{cl}-x~sin\theta_{cl}
sin\varphi_{cl}-z~cos\theta_{cl}
sin\varphi_{cl}+y~cos\varphi_{cl}}{R_{cl}~cos\varphi_{cl}-x~sin\theta_{cl}
sin\varphi_{cl}-z~cos\theta_{cl} sin\varphi_{cl}-y~sin\varphi_{cl}}\right)\nonumber\\
\end{eqnarray*}
and
$$r=\sqrt{r_{cl}^{2}+|{\bf r'}|^{2}-2 x r_{cl}}$$
In analogy with what done in Sect. \ref{pm_sec}, we calculate the
projection of the acceleration felt by the star (eq. \ref{acc_eq}) on ${\bf
r'}$. 
The internal acceleration is always given by eq. \ref{pmc_eq}.

The term associated to the external field can be calculated separately for the
three galactic components
\begin{eqnarray}
\frac{{\bf r'}}{|{\bf r'}|} \cdot ({\bf r'} \cdot {\bf \nabla}) \nabla
\phi_{b}&=&\frac{G M_{b} |{\bf
r'}|}{r_{cl}(r_{cl}+c)^{3}}[(r_{cl}+c)(\tilde{y}^{2}+\tilde{z}^{2})-2 r_{cl}
\tilde{x}^{2}]\nonumber\\
\frac{{\bf r'}}{|{\bf r'}|} \cdot ({\bf r'} \cdot {\bf \nabla}) \nabla
\phi_{d}&=&G
M_{d} |{\bf r'}|\left\{\frac{1+\frac{a
b^{2}}{(b^{2}+Z_{cl}^{2})^{\frac{3}{2}}}(\tilde{x}^{2}cos^{2}\theta+\tilde{z}^{2}sin^{2}\theta)}
{\left[R_{cl}^{2}+\left(a+\sqrt{b^{2}+Z_{cl}^{2}}\right)^{2}\right]^{\frac{3}{2}}}-
3\frac{\left(r_{cl}^{2}+\frac{2 a
Z_{cl}^{2}}{\sqrt{b^{2}+Z_{cl}^{2}}}\right) \tilde{x}^{2}+\frac{a^{2}
Z_{cl}^{2}}{b^{2}+Z_{cl}^{2}}(\tilde{x}^{2}cos^{2}\theta+\tilde{z}^{2}sin^{2}\theta)}
{\left[R_{cl}^{2}+\left(a+\sqrt{b^{2}+Z_{cl}^{2}}\right)^{2}\right]^{\frac{5}{2}}}\right\}\nonumber\\
\frac{{\bf r'}}{|{\bf r'}|} \cdot ({\bf r'} \cdot {\bf \nabla}) \nabla
\phi_{h}&=&\frac{2
v_{0}^{2} |{\bf
r'}|}{(d^{2}+r_{cl}^{2})^{2}}[(d^{2}+r_{cl}^{2})(\tilde{y}^{2}+\tilde{z}^{2})+(d^{2}-r_{cl}^{2})\tilde{x}^{2}].
\label{complf_eq}
\end{eqnarray}
where
$$\tilde{x_{i}}=\frac{x_{i}}{|{\bf r'}|}$$
The acceleration due to the angular motion of the star is given by eq.
\ref{pmat_eq}, where 
\begin{eqnarray*}
{\bf \Omega}&=&-\frac{L_{z}}{R_{cl}^{2}}cos\theta_{cl}{\bf \hat{i}}-\dot{\theta}_{cl}{\bf
\hat{j}}+\frac{L_{z}}{R_{cl}^{2}}sin\theta_{cl}{\bf \hat{k}}\nonumber\\
\dot{\theta}_{cl}&=&\frac{Z_{cl}\dot{R}_{cl}-R_{cl}\dot{Z}_{cl}}{r_{cl}^{2}}\nonumber
\end{eqnarray*}
and $L_{z}$ is the z-component of the cluster angular momentum. 
As in the case of the point-mass potential, only the Coriolis ({\bf $a_{cor}$}) 
and the centrifugal+tidal ({\bf $a_{ct}$}) terms of the
acceleration have a projection on ${\bf r'}$ different from zero. So
\begin{eqnarray}
\frac{{\bf r'}}{|{\bf r'}|} \cdot {\bf
a_{\Omega}}&=&\frac{{\bf r'}}{|{\bf r'}|} \cdot ({\bf
a_{ct}} + {\bf
a_{cor}})\nonumber\\
&=&-\frac{L_{z}^{2} |{\bf r'}|}{R_{cl}^{4}
}[(\tilde{x}~sin\theta_{cl}+\tilde{z}~cos\theta_{cl})^{2}+\tilde{y}^{2}]-
\dot{\theta}_{cl}^{2} |{\bf r'}|(\tilde{x}^{2}+\tilde{z}^{2})+
\frac{2 L_{z} \dot{\theta}_{cl} |{\bf r'}|}{R_{cl}^{2}} \tilde{y} 
(\tilde{x}~cos\theta_{cl}-\tilde{z}~sin\theta_{cl})\nonumber\\
& &-2\dot{\theta}_{cl}(\tilde{x}~v_{z}-\tilde{z}~v_{x})-\frac{2 L_{z}}{R_{cl}^{2}}[\tilde{x}~v_{y}sin\theta_{cl}-\tilde{y}
(v_{x}sin\theta_{cl}+v_{z}cos\theta_{cl})+\tilde{z}~v_{y}cos\theta_{cl}].
\label{compla_eq}
\end{eqnarray}

It is convenient to introduce the characteristic densities
\begin{eqnarray}
\Gamma&=&\sum_{i}\Gamma_{i}\nonumber\\ 
\Gamma_{i}&=&-\frac{{\bf r'} \cdot {\bf a_{i}}}{G |{\bf
r'}|^{2}}\nonumber\\
\Gamma_{cor}&=&-\frac{2 L}{G r_{cl}^{2}\sqrt{\tilde{x}^{2}+\tilde{y}^{2}}|{\bf
r'}|}\times\nonumber\\
& &\left\{\dot{\theta}_{cl}(\tilde{x}~cos(2\pi\eta_{4})+\tilde{y}\tilde{z}~
sin(2\pi\eta_{4}))+\frac{L_{z}}{R_{cl}^{2}}\left[(\tilde{x}^{2}+\tilde{y}^{2})
sin(2\pi\eta_{4})sin\theta_{cl}+(\tilde{x}\tilde{z}sin(2\pi\eta_{4})-\tilde{y}
cos(2\pi\eta_{4}))cos\theta_{cl}\right]\right\}.
\label{complb_eq}
\end{eqnarray}
where ${\bf a_{i}}$ is either the acceleration due to the i-th galactic component
(${\bf a_{i}}={\bf (r' \cdot \nabla)\nabla \phi_{i}}$) or the
centrifugal+tidal acceleration, $L$ is the angular momentum of the superstar and
$\eta_{4}$ is a random number uniformly distributed between 0 and 1 (see eq.
\ref{vel_eq}).
It is therefore possible to combine eq.s \ref{pmc_eq}, \ref{complf_eq}, \ref{compla_eq} and
\ref{complb_eq} to obtain the projection of the total 
acceleration felt by the star on ${\bf r'}$
$${\bf \nabla \phi_{eff}}=\frac{G M_{cl}}{|{\bf r'}|^{2}}-G \Gamma 
|{\bf r'}| -\frac{G \Gamma_{cor} r_{cl^{2}}}{|{\bf r'}|}$$

The tidal radius will therefore be
$$r_{t}=\left(-\frac{q}{2}+\sqrt{\frac{q^{2}}{4}+\frac{p^{3}}{27}}\right)^{\frac{1}{3}}+
\left(-\frac{q}{2}-\sqrt{\frac{q^{2}}{4}+\frac{p^{3}}{27}}\right)^{\frac{1}{3}}$$
with
\begin{eqnarray*}
p&=&\frac{\Gamma_{cor} r_{cl}^{2}}{\Gamma}\nonumber\\
q&=&-\frac{M_{cl}}{\Gamma}.
\end{eqnarray*}
Again, the Coriolis term has null mean over a large number of random extractions
of $\eta_{4}$, so we can calculate $r_{t}$ assuming $\Gamma_{cor}=0$.
$$r_{t}=\left(\frac{M_{cl}}{\Gamma}\right)^{\frac{1}{3}}$$
Note that at the Lagrangian point $(\tilde{x},\tilde{y},\tilde{z})=(1,0,0)$ 
the above formula is equivalent to the estimate provided by Allen et al.
(2006).

The effective potential will be
$$\phi_{eff}(|{\bf r'}|)=\phi_{cl}(|{\bf r'}|)-\frac{G \Gamma |{\bf
r'}|^{2}}{2}
$$
In $r_{t}$ the above quantity is
$$\phi_{eff}(r_{t})=-\frac{3 G (M_{cl}^{2} \Gamma)^{\frac{1}{3}}}{2}$$

\label{lastpage}

\end{document}